\documentclass[10pt,twocolumn,twoside,journal]{IEEEtran}
\IEEEoverridecommandlockouts
\usepackage{subfigure} 
\usepackage{graphicx}

\usepackage{amsmath,graphicx,amssymb,mathtools,bm}
\usepackage{subfigure}
\usepackage{hyperref}
\usepackage{cite}
\usepackage{amsmath,amssymb,amsfonts}
\usepackage{algorithmic}
\usepackage{graphicx}
\usepackage{textcomp}
\usepackage{xcolor}
\usepackage{verbatim}  
\usepackage{graphicx}  
\usepackage{bm}  
\usepackage{mathrsfs} 
\usepackage{algorithm} 
\usepackage{algorithmic} 
\usepackage{booktabs}
\usepackage{textcomp}  
\usepackage{multirow}  
\usepackage{lettrine}   
\usepackage{graphicx}  
\usepackage{color}  
\usepackage{amsmath}
\usepackage{amssymb}

\def\BibTeX{{\rm B\kern-.05em{\sc i\kern-.025em b}\kern-.08em
    T\kern-.1667em\lower.7ex\hbox{E}\kern-.125emX}}
\begin{document}
	\newcommand{\tabincell}[2]{\begin{tabular}{@{}#1@{}}#2\end{tabular}} 

\title{ An Off-grid Compressive Sensing Strategy for the Subarray Synthesis  of Non-uniform Linear Arrays
}

\author{Songjie Yang,
	Wanting Lyu,
        Zhongpei Zhang,~\IEEEmembership{Member,~IEEE},

\thanks{This work was supported in part by the National Key Research and Development Program of China under Grant 2020YFB1806800. (\textit{Corresponding author:
Zhongpei~Zhang}.)
}

\thanks{Songjie Yang, Wanting Lyu and Zhongpei Zhang are with the National Key Laboratory of Science and Technology on Communications, University of Electronic Science and Technology of China, Chengdu 611731, China. (e-mail:
	yangsongjie@std.uestc.edu.cn; lyuwanting@yeah.net;
	zhangzp@uestc.edu.cn).}
}
\maketitle

\begin{abstract}
With the increasing popularity of large-scale antenna arrays, the subarraying technology becomes more attractive.
In this paper, we propose two effective subarraying methods right after formulating the subarray synthesis as a compressive sensing (CS) problem: i) Orthogonal matching pursuit based subarray synthesis (OMP-SS), a common CS approach which can be used for the subarray synthesis to attain the subarray information (the subarray number, the number of elements per subarray and corresponding excitation coeffcients) and ii) Off-grid orthogonal matching pursuit based subarray synthesis (OGOMP-SS), an advanced approach for optimizing antenna elements positions and the subarray information mentioned above simultaneously. In addition, two user-defined modes are designed for different application scenarios, wherein, mode-1 is to optimize the pattern synthesis performance for the given the number of subarrays, and mode-2 is to obtain the minimum number of subarrays for the cases when the pattern synthsis accuracy is satisfied. Finally, our simulation results reveal that it is of paramount significance to optimize antenna elements positions for the subarray synthesis performance on the one hand and demonstrate the excellent performances of proposed schemes in comparison with other competitive state-of-the-art subarray synthesis methods on the other hand.

\end{abstract}
\section{Introduction}
\lettrine[lines=2]{L}arge-scale antenna arrays for beam pattern synthesis are widely adopted in scenarios where specific pattern information such as shaped pattern and sidelobe levels (SLL) need to be specified. So far, large scale antenna arrays have been used in a variety of scenarios \cite{sc1, sc2, sc3, sc4,sc5,sc6,sc7}, e.g.,  radar systems, 5G millimeter wave communications and satellite communications. For the conventional array architecture, each antenna element is excited by a control point, inducing superior performance but unacceptable cost and energy. Therefore, it is of paramount significance to reach the best compromise between performance and cost. Faced with the trade-off problem, the subarraying technology is proposed to reconstruct a clustered architecture based on the subarray synthesis, in which the $N$ antenna elements in the array are grouped in $K$ ($K$ $\ll N$) subarrays, where each subarry rather than each antenna element is excited by a control point, shrinking cost and energy by a large margin. However, the subarrarying technology, which needs to determine the layout of subarrays (the subarray number,
the subarray shape, the number of elements per subarray) and the excitation coefficient of each subarry, is a NP-hard problem. Therefore, it is very important to propose effective subarray synthesis methods.

\begin{figure}
	\centering
	\includegraphics[width=0.475\textwidth]{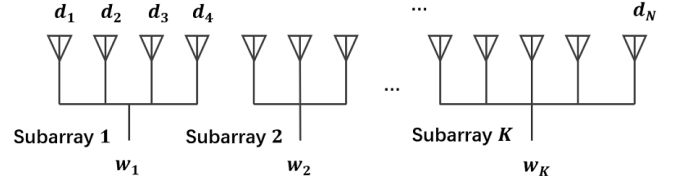}
	\caption{Geometry of the uniform linear array with N elements into $K$ nonoverlapped subarrays.}
\end{figure}

In the previous researches on subarraying, there emerged different effective approaches to the subarray synthesis. 
\cite{ss1} applied the genetic algorithm (GA) optimization  process to  optimize the subarray partition functions jointly with the weights for the given subarray number. \cite{ss2} proposed a subarraying method with an efficient hybrid differential evolution algorithm, which makes use of hybrid chromosomes (constituted by real and integer genes) and does not need any coding of the real variables. In \cite{ss3}, a hybrid GA was proposed to optimize the subarray size and corresponding excitation coefficients to minimize the maximum SLL. However, these heuristic algorithm based schemes are usually time-consuming when dealing with large arrays. Unlike the above schemes, \cite{ss4} considered that the subarraying can be solved by exploiting the correlation between  difference excitations and independently optimum sum and regarded the subarraying problem as searching the optimal path in an incomplete binary tree performed by a swapping strategy, which is called tree-searching method. Moreover, the contiguous partition method (CPM) was proposed in \cite{ss5}, where the clustering of the antenna elements into contiguous subarrays and the solution of the excitation coefficients were simultaneously carried out by an unconstrained excitation matching strategy which reformulated the clustering problem into searching the optimal path in an incomplete binary tree.

More recently, several Compressive Sensing (CS) based approaches to subarraying have been proposed. The CS based method---the sparse-regularized
solver was first proposed in \cite{sub-cs1}: therein, the subarray synthesis problem was recast as a sparse signal reconstruction problem on the basis of the user-defined dictionary. In addition, this scheme can be seamlessly applied also to overlapped arrangements depending on the user-defined dictionary. According to the CS theory, \cite{sub-cs2} adopted a convex programming method to attain the corresponding parameters including the number of subarrays and the subarray weights and sizes can be determined simultaneously to satisfy the prescribed focused and/or shaped beam patterns. In \cite{sub-cs3}, a novel method for the synthesis of physically-contiguous clustered linear arrays is proposed by exploiting an innovative Total-Variation Compressive Sensing (TV-CS) framework. However, all the above CS based schemes lack the optimization of antenna elements positions, resulting in less degrees of freedom. 
On the one hand, in the common array synthesis problem, since the correlation between the array factor and element positions is strongly nonlinear, convex programming methods, widely used in fully-populated tapered designs, is here hindered \cite{GO1}. In order to effectively cope with the non-linear nature of the problem, a variety of global optimization strategies have been proposed, e.g., genetic algorithm (GA)\cite{GO2}, simulated annealing (SA) \cite{GO3} and biogeography-based techniques \cite{GO4}. On the other hand, these global optimization strategies may also play a role in the subarray synthesis problem.
 Lately, a hybrid optimization scheme taking into account the positions of the antenna elements was documented in \cite{sub-cs4}, where convex programming and particle swarm optimization (PSO) were adopted for the synthesis of excitations and locations of nonuniformly spaced linear subarrays to minimize the subarray number. It is notewothy that heuristic algorithms such as PSO have high computational complexity, especially in the large scale antenna array.

For the purpose of improving the performances of the subarray synthesis while taking into account the complexity, we propose a novel subarray synthesis scheme on the basis of CS, which optimizes all the parameters mentioned in the above scheme simultaneously, i.e., we also take into accout the positions of antenna elements as well as \cite{sub-cs4} does. On the basis of the CS framework of the subarray synthesis, we extend it to a more flexible model involving the elements positions optimization. 
Viewed from another perspective, our proposed scheme for the subarray synthesis can be regarded as an off-grid CS problem, which not only recovers the sparse vector information (the subarray number, the number of elements per subarry and the excitation coefficients per subarry), but also further optimize the dictionary matrix information (antenna elements positions). Up to this point, CS based schemes in \cite{sub-cs1,sub-cs2,sub-cs3} can be recast as on-grid CS frameworks, which ignore the optimization of the dictionary matrix. To solve this CS problem, Orthogonal Matching Pursuit (OMP) is adopted for the subarray synthesis by recovering the sparse vector information, and then an improved off-grid OMP approach is proposed to further optimize the dictionary matrix information. In addition, we design two user-defined modes: i) required sparsity (i.e., the number of power control points) and ii) required pattern matching error between the desired pattern and the archieved pattern. For concreteness, these two modes may be used in different scenarios. The former aims to attain the optimal performance (pattern matching error) when the number of power control points in the physical architecture designed in advance is known, the latter is to minimize the number of power control points while satisfying the required pattern matching error.    
 
Our major contributions are summarized as follows:
\begin{itemize}
	\item We first fomulate the subarray synthesis problem as a CS framework by using a sparse basis matrix to sparsely express the excitation coefficients vector.

	\item Then we propose OMP algorithm based subarray synthesis (OMP-SS) to attain the subarray information via solving the CS problem. In addition, two user-defined modes given different requirements are considered to design the subarray synthesis algorithm.
	
	\item We further propose an off-grid OMP based subarray synthesis (OGOMP-SS) to optimize the antenna elements positions and excitation coefficients simultaneously, which uses the off-grid thought to solve the basis mismatch problem to optimize antenna elements positions.  
	
	\item Finally, numerical experiments are carried out to demonstrate the effectiveness of our proposed schemes and compare them with competitive state-of-the-art subarray synthesis methods.  
\end{itemize}

The remainder of this article is organized as follows. In Section \S\ref{PF}, we formulate the subarray synthesis as the on-grid CS framework, a common modeling of other researches on CS based subarraying. After that, an off-grid CS framework based subarray synthesis is proposed in \S\ref{off-grid}.  Several numerical simulation results are then presented in Section \S\ref{SR} to validate the effectiveness of the proposed scheme, as well as in comparison with the state-of-the-art schemes. Finally, some conclusions and final remarks are documented in Section \S\ref{Con}. 
\begin{figure}
	\centering
	\includegraphics[width=0.475\textwidth]{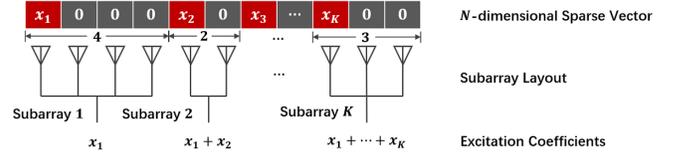}
	\caption{Array information covered by the $N$-dimensional sparse vector.}
\end{figure}
\section{Problem Formulation For The Subarray Synthesis}\label{PF}

\figurename{1} exhibits a linear array of $N$ uniformly displaced antenna elements partitioned into $K$ nonoverlapped subarrays with unequal
weights and sizes. Suppose that the set of antenna elements locations is $\bm{d}=\{d_1,d_2,\cdots,d_n\}$ and the corresponding excitation coefficients are $\mathbf{w}=\{w_1,w_2,\cdots,w_n\}$, The array factor function is given by
\begin{equation}\label{F1}
	F(\bm{\theta})=\sum_{i=1}^{N}w_i e^{{j\frac{2\pi}{\lambda}d_isin(\bm{\theta})}}
\end{equation}
where $N$ is the number of antenna elements, $\lambda$ is the wavelength, $\bm{\theta}=\{\theta_1,\theta_2,\cdots,\theta_M\}$ is the set of observed angles, $w_i$ is the excitation coefficient of the $i$-th element and $d_i$ denotes the $i$-th element location.

(\ref{F1}) can be rewritten as
\begin{equation}\label{F2}
	F(\bm{\theta})=\mathbf{\Phi w}
\end{equation}
where 
\begin{equation}
\mathbf{\Phi}=
	\begin{bmatrix}
	e^{{j\frac{2\pi}{\lambda}d_1sin(\theta_1)}} & e^{{j\frac{2\pi}{\lambda}d_2sin(\theta_1)}} & \cdots & e^{{j\frac{2\pi}{\lambda}d_Nsin(\theta_1)}} \\
	e^{{j\frac{2\pi}{\lambda}d_1sin(\theta_2)}} & e^{{j\frac{2\pi}{\lambda}d_2sin(\theta_2)}} & \cdots & e^{{j\frac{2\pi}{\lambda}d_Nsin(\theta_2)}} \\
	\vdots & \vdots & \ddots & \vdots \\
	e^{{j\frac{2\pi}{\lambda}d_1sin(\theta_M)}} & e^{{j\frac{2\pi}{\lambda}d_2sin(\theta_M)}} & \cdots & e^{{j\frac{2\pi}{\lambda}d_Nsin(\theta_M)}}
\end{bmatrix}
\end{equation}
and $\mathbf{w}=[w_1,w_2,\cdots,w_N]^T$.

To formulate (\ref{F2}) as a CS expression, a sparse basis matrix $\bm{\Psi}$ is supposed to express $\mathbf{w}$ sparsely. That is,
\begin{equation}
	F(\bm{\theta})=\bm{\Phi\Psi} \mathbf{x}
\end{equation}
where
\begin{equation}
	\bm{\Psi}=
	\begin{bmatrix}
		1 & 0 & \cdots & 0 \\
		1 & 1 & \cdots & 0 \\
		\vdots & \vdots & \ddots & \vdots \\
		1 & 1 & \cdots & 1
	\end{bmatrix}_{N\times N}
\end{equation}
is a binary sparse basis matrix, and $\mathbf{x}$ is represented by 
\begin{equation}
	\mathbf{x}=[w_1,w_2-w_1,\cdots,w_n-w_{n-1}]^T
\end{equation}

In particular, the information of the subarray synthesis can be observed from the reconstructed sparse vector $\mathbf{x}$. As shown in the \figurename{2}, the number of non-zero elements represents the number of subarrays, the non-zero value is related to the excitation coefficient of the corresponding subarray, and the spacing between non-zero elements denotes the subarray sizes.

For the flexibility of the subarray synthesis, two user-defined modes are designed suitable for different scenarios are designed as follows.

The subarray synthesis problem with user-defined mode-1, aiming to fit the desired radiation pattern as much as possible given the sparisty is fixed, can be represented as
\begin{equation}\label{Fm1}
\begin{aligned}
&\underset{\mathbf{x}}{\rm arg \ min} \ \Vert \bar{F}(\bm{\theta})-\mathbf{Ax}\Vert_2 \\
& {\rm s.t.} \ \Vert \mathbf{x}\Vert_0=K
\end{aligned}
\end{equation}
where $\bar{F}(\bm{\theta})\in\mathbb{C}^M$ is the desired pattern and $\mathbf{A}=\bm{\Phi\Psi}$.

Another user-defined mode for the subarray synthesis problem is described as minimizing the non-zero number of $\mathbf{x}$ while satisfying the matching accuracy and given by
\begin{equation}\label{F}
	\begin{aligned}
		&  \underset{\mathbf{x}}{\rm arg \ min} \ \Vert \mathbf{x}\Vert_0\\
		&  {\rm s.t.} \ \Vert \bar{F}(\bm{\theta})-\mathbf{Ax}\Vert_2 \leq \varepsilon
	\end{aligned}
\end{equation}
where $\varepsilon$ is the threshold controlling the matching accuracy.
\section{An Improved Off-Grid OMP Based Subarray Synthesis}\label{off-grid}
\subsection{OMP-SS}
For the sparse reconstruction problem of the subarray synthesis, OMP is adopted due to its ease of implementation and its speed \cite{OMP}, which is a greedy CS recovery algorithm that chooses the best fitting column of the dictionary matrix in each iteration. After that, the least squares algorithm is carried out in the subspace spanned by all collected columns. 
In the light of the process of standard OMP under the assumption of sparsity of $K$ or iteration accuracy $\varepsilon$ illustrated, shown in Algorithm \ref{on-omp}, the sparse vector and  the corresponding subarray information can be easily attained. 

\begin{algorithm}[!t] 
	\caption{OMP-SS For Two User-defined Modes } 
	\label{on-omp}      
	\begin{algorithmic}[1] 
		\footnotesize{
			\REQUIRE {$\bm{\Phi}$,$\bm{\Psi}$,$\bar{F}(\bm{\theta})$ and $K$ or $\varepsilon$. }
			\ENSURE {Optimal $\mathbf{x}^\prime$.} 
			\STATE{
			\emph{User-defined mode-1.}
		}
			\STATE{$\textbf{Initialize:}$ Resdual $\bm{r}^0=\bar{F}(\bm{\theta})$, solution $\mathbf{x}^0=0$, solution support $\Lambda^0=\emptyset$ and t=0.
			}

			\REPEAT
			\STATE{ Match: $\mathbf{z}^t=(\bm{\Phi\Psi})^H\bm{r}^t$.}
				\STATE{Update Support: $\Lambda^{t+1}=\Lambda^t\cup\{{\underset{i}{\rm arg \ max}}\vert \mathbf
					{z}^t(i)\vert\}$.}
				\STATE{Update Residual:
					$\mathbf{x}^{t+1}=\underset{\mathbf{f}:supp(\mathbf{f}\subseteq\Lambda^t)}{\rm arg \ min} \Vert\mathbf{z}-\bm{\Phi\Psi\mathbf{f}}\Vert_2$ \\
					\ \ \ \ \ \ \ \ \ \ \ \ \ \ \ \ \ \ \ \ \ $\mathbf{z}^{t+1}=\mathbf{z}-\bm{\Phi\Psi}\mathbf{x}^{t+1}$ \\
						\ \ \ \ \ \ \ \ \ \ \ \ \ \ \ \ \ \ \ \ \
					$t=t+1$.
				}
			\UNTIL{$t\geq K$.}
			\STATE { $\mathbf{x}^\prime\leftarrow\mathbf{x}^t$.}
			
				\STATE{
				\emph{User-defined mode-2.}
				\STATE{ Same as lines 2 to 8, except that the iteration termination condition in line 7 is changed to $\Vert\bar{F}(\bm{\theta})-\bm{\Phi\Psi}\mathbf{x}^t\Vert_2\leq\varepsilon$
				.}
			}
		}
	\end{algorithmic}
\end{algorithm}

\subsection{OGOMP-SS}
To further improve the performance of subarray synthesis, the antenna elements positions optimization is taken into account, which is equivalent to the dictionary optimization problem in the off-grid CS. In fact, this is the so-called basis mismatch problem \cite{mismatch}: no matter how to finely grid the parameter space the sources  may not be located in the center of the grid cells and there exists mismatch between the actual and the assumed bases for sparsity as a consequence. To put it crudely, the initial antenna elements positions are not necessarily optimal for the sparse reconstruction of $\mathbf{x}$. Thus, we propose to iterately reconstruct $\mathbf{x}$ and optimize $\bm{\Phi}$ contained $N$ elements positions information to improve the subarray synthesis performance.

First considering the subarray synthesis for user-defined mode-1, for the purpose of optimizing $\bm{d}$ and $\mathbf{x}$ simultaneously, (\ref{Fm1}) can be expressed as 
\begin{equation}\label{Fm2}
	\underset{\bm{d},\mathbf{x}}{\rm arg \ min} \ \Vert \bar{F}(\bm{\theta})-\bm{\Phi(d)\Psi}\mathbf{x}\Vert_2 
\end{equation}

The vector $\mathbf{x}$ can be sparsely reconstructed in line with Algorithm \ref{on-omp} given $\bm{d}$ is known. Generally, the initial $\bm{d}$ is an increasing sequence of the antenna element spacing $\lambda/2$. That is to say, the first step of OGOMP-SS is to attain the excitation coefficients at the initial elements positions via OMP-SS.
Then we propose a perturbation based strategy to optimize $\bm{d}$.

\begin{algorithm}[!t] 
	\caption{OGOMP-SS For Two User-defined Modes } 
	\label{off-omp}      
	\begin{algorithmic}[1] 
		
		\REQUIRE { $\bm{\Phi}$, $\bm{\Psi}$, $\bar{F}(\bm{\theta})$ and $K$ or $\varepsilon$.  }
		\ENSURE {Optimal $\bm{d}^\prime$ and $\mathbf{x}^\prime$.} 
		\footnotesize{
			\STATE{\emph{User-defined mode-1}.	
			}
			\STATE{$\textbf{Initialize:}$ Attain the $K$-dimensional  nonzero vector $\hat{\mathbf{x}}$ and corresponding basis matrix $\hat{\bm{\Psi}}$ via Algorithm \ref{on-omp} given the sparsity is $K$.
			}
			\REPEAT
			\STATE{Denote $\mathbf{w}=\hat{\bm{\Psi}}\hat{\mathbf{x}}$ and $\bm{y}=\bar{F}(\bm{\theta})-\sum_{i=1}^{N} w_i\bm{\Phi}({d}_{i})$.	
			}
			\STATE{ Compute $\bm{\eta}$ in the light of (\ref{G}) and (\ref{eta}).
			}
			\STATE{ Update $\bm{d}$ and $\bm{\Phi}(\bm{d})$ by $d_i\leftarrow d_i+\eta_id_i$.
			}
			\STATE{ Obtain $\hat{\mathbf{x}}$ via 	$\hat{\mathbf{x}}=((\bm{\Phi}^\prime\hat{\bm{\Psi}})^H\bm{\Phi}^\prime\hat{\bm{\Psi}})^{-1}(\bm{\Phi}^\prime\hat{\bm{\Psi}})^H \bar{F}(\bm{\theta})$.

			}
			\UNTIL{\emph{Q} iterations are reached.}
			\STATE { $\bm{d}^\prime\leftarrow\bm{d}$, $\mathbf{x}^\prime\leftarrow\hat{\mathbf{x}}$.}
		}
		\STATE{\emph{User-defined mode-2}.	
		}
		\STATE{$\textbf{Initialize:}$ $h=1$.
		}
		\REPEAT
		\STATE{ Attain the $h$-dimensional  nonzero vector $\hat{\mathbf{x}}$ and corresponding basis matrix $\hat{\bm{\Psi}}$ via Algorithm \ref{on-omp} given the sparsity is $h$.
		}
		\REPEAT
		\STATE{Same as lines 4-7 above.}
		\UNTIL{$\varepsilon$ is satisfied or \emph{Q} iterations are reached.}
		\STATE{$h=h+1$.}
		\UNTIL{$\varepsilon$ is satisfied.}
		\STATE { $\bm{d}^\prime\leftarrow\bm{d}$, $\mathbf{x}^\prime\leftarrow\hat{\mathbf{x}}$.}
	\end{algorithmic}
\end{algorithm}
Suppose that $\hat{\mathbf{x}}\in\mathbb{C}^{K}$ is the $K$ nonzero elements of the reconstructed vector $\mathbf{x}$ and $\hat{\bm{\Psi}}\in\mathbb{C}^{N\times K}$ denotes the corresponding $K$ columns derived from $\bm{\Psi}$. (\ref{Fm2}) can be rewritten as 
\begin{equation}\label{Fm3}
	\begin{aligned}
		&\underset{\bm{d}}{\rm arg \ min} \ \Vert \bar{F}(\bm{\theta})-\bm{\Phi}(\bm{d})\hat{\bm{\Psi}}\hat{\mathbf{x}}\Vert_2 \\
		=&\underset{\bm{d}}{\rm arg \ min} \ \Vert \bar{F}(\bm{\theta})-\bm{\Phi}(\bm{d})\mathbf{w}\Vert_2 \\
		=&\underset{\bm{d}}{\rm arg \ min} \ \Vert \bar{F}(\bm{\theta})-\sum_{i=1}^{N}
		w_i\bm{\Phi}({d}_{i})\Vert_2 \\
		=&\underset{\bm{\eta}}{\rm arg \ min} \ \Vert \bar{F}(\bm{\theta})-\sum_{i=1}^{N}
		w_i\bm{\Phi}({d}_{i}+\eta_i {d}_i)\Vert_2
	\end{aligned}
\end{equation}
where $\bm{\Phi}(d_i)$ denote the $i$-th column of $\bm{\Phi}$ and $\eta d_i$ denotes the perturbation of the $i$-th element location.

To cope with (\ref{Fm3}), the first order Taylor expansion can be used to approximate $\bm{\Phi}({d}_{i}+\eta_i {d}_i)$, where
\begin{equation}
\bm{\Phi}({d}_{i}+\eta_i {d}_i)\approx \bm{\Phi}(d_i) + \eta_i d_i\frac{\partial \bm{\Phi}(d_i)}{\partial d_i}
\end{equation}
after that, (\ref{Fm3}) is simplified to
\begin{equation}\label{Fm4}
	\begin{aligned}
			&\underset{\bm{\eta}}{\rm arg \ min} \ \Vert \bar{F}(\bm{\theta})-\sum_{i=1}^{N}
		w_i\bm{\Phi}({d}_{i}) - \sum_{i=1}^{N} \eta_i w_id_i\frac{\partial \bm{\Phi}(d_i)}{\partial d_i} \Vert_2 \\
		=&\underset{\bm{\eta}}{\rm arg \ min} \ \Vert \bm{y}-\bm{G} \bm{\eta} \Vert_2
	\end{aligned}
\end{equation}
where 
\begin{equation}\label{G}
\bm{G}=\begin{bmatrix}
	w_1d_1 \frac{\partial e^{j\frac{2\pi}{\lambda}d_1sin(\theta_1) }}{\partial d_1} & \cdots & 	w_Nd_N \frac{\partial e^{j\frac{2\pi}{\lambda}d_Nsin(\theta_1) }}{\partial d_N}  \\
	\vdots  & \ddots & \vdots \\
	w_1d_1 \frac{\partial e^{j\frac{2\pi}{\lambda}d_1sin(\theta_M) }}{\partial d_1}  & \cdots & w_Nd_N \frac{\partial e^{j\frac{2\pi}{\lambda}d_Nsin(\theta_M) }}{\partial d_N}
\end{bmatrix}
\end{equation}
 and $\bm{y}=\bar{F}(\bm{\theta})-\sum_{i=1}^{N} w_i\bm{\Phi}({d}_{i})$.

It is noteworthy that (\ref{Fm4}) is a standard least squares expression, and its closed-form solution is given by
\begin{equation}\label{eta}
	\bm{\eta}=(\bm{G}^H\bm{G})^{-1}\bm{G}^H\bm{y}
\end{equation}

To this end, the dictionary matrix is updated by $d^\prime_i=d_i+\eta_id_i$ and recorded as $\bm{\Phi}^\prime$. Furthermore, the sparse solution of $\mathbf{x}$ is updated by (\ref{Fm3}) and is written as
\begin{equation}\label{x}
	\mathbf{x}^\prime=((\bm{\Phi}^\prime\hat{\bm{\Psi}})^H\bm{\Phi}^\prime\hat{\bm{\Psi}})^{-1}(\bm{\Phi}^\prime\hat{\bm{\Psi}})^H \bar{F}(\bm{\theta})
\end{equation}

 Up to this point, the excitation coefficients can be attained by $\mathbf{w}=\hat{\bm{\Psi}}\mathbf{x}^\prime$. Of course, new $\bm{d}^\prime$ can be attained by the proposed algorithm via $\mathbf{x}^\prime$. Thus, $\bm{d}$ and $\mathbf{x}$ can be iterately optimized until the stop condition is met. 
 
The subarray synthesis for user-defined mode-2 of OGOMP-SS can be solved by setting the initial sparsity $h=1$ followed by optimizing elements positions and excitation coefficients, while determining whether it meets the user-defined matching accuracy. If not, let $h=h+1$ and repeat the above steps. Then the minimum number of subarrays $h$ satisfying the matching accuracy can be found iteratively in this manner. 
In summary, the whole synthesis flow is shown in Algorithm \ref{off-omp}.

 \begin{table*}
 	\caption{Numerical Validation of different schemes for different desired patterns with user-defined mode-2. }
 	\label{all}
 	\centering
 	\begin{tabular}{ |c|c|| c | c |c|c| }
 		\hline
 		Desired Pattern&\tabincell{c}{User-defined\\ $\xi$} & Scheme in \cite{sub-cs3} & Scheme in \cite{sub-cs4} &Proposed OMP-SS&Proposed OGOMP-SS \\ \hline\hline
 		\multirow{6}{*}{\tabincell{c}{Dolph-Tschebyscheff \\($N$=20, SLL=-20 dB)}} &  1.0$\times10^{-2}$ & $\chi\approx$0.35 &--&$\chi$=0.35, SLL=-17.86 &$\chi$=0.05, SLL=-18.65  \\ \cline{2-6}
 		& 1.0$\times 10^{-3}$& $\chi\approx$0.64, SLL$\approx$-17&--&$\chi$=0.65, SLL=-19.09  &$\chi$=0.15, SLL=-19.64 \\
 		\cline{2-6}
 		& 1.0$\times10^{-4}$ & $\chi\approx$0.80&--&$\chi$=0.85, SLL=-19.91 &$\chi$=0.25, SLL=-19.76 \\
 		\cline{2-6}
 		& 2.8$\times 10^{-3}$ & --&$\chi$=0.05, SLL=-18.41&$\chi$=0.55, SLL=-19.81&$\chi$=0.05, SLL=-18.65 \\
 		\cline{2-6}
 		& 7.3$\times 10^{-4}$ & --&$\chi$=0.15, SLL=-18.95&$\chi$=0.75, SLL=-19.47&$\chi$=0.15, SLL=-19.64 \\
 		\cline{2-6}
 		& 1.7$\times 10^{-4}$ & --&$\chi$=0.25, SLL=-19.91&$\chi$=0.85, SLL=-19.83&$\chi$=0.25, SLL=-19.76 \\
 		
 		\hline
 		\multirow{6}{*}{\tabincell{c}{Dolph-Tschebyscheff \\($N$=100, SLL=-30 dB)}} & 1.0$\times 10^{-2}$ & $\chi\approx$0.25, SLL$\approx$-20&--&$\chi$=0.09, SLL=-24.31&$\chi$=0.07, SLL=-25.88 \\ \cline{2-6}
 		& 1.0$\times 10^{-3}$ & $\chi\approx$0.49, SLL$\approx$-27&--&$\chi$=0.25, SLL=-28.34&$\chi$=0.09, SLL=-29.83 \\
 		\cline{2-6}
 		& 1.0$\times 10^{-4}$& $\chi\approx$0.73, SLL$\approx$-29&--&$\chi$=0.67, SLL=-29.40&$\chi$=0.09, SLL=-29.83 \\
 		\cline{2-6}
 		& 2.1$\times 10^{-3}$ & --&$\chi$=0.13, SLL=-26.70&$\chi$=0.17, SLL=-27.97&$\chi$=0.09, SLL=-29.83 \\
 		\cline{2-6}
 		& 5.3$\times 10^{-4}$ & --&$\chi$=0.21, SLL=-29.67&$\chi$=0.35, SLL=-28.74&$\chi$=0.09, SLL=-29.83 \\
 		\cline{2-6}
 		& 8.4$\times 10^{-5}$ & --&$\chi$=0.39, SLL=-29.52&$\chi$=0.71, SLL=-29.28&$\chi$=0.09, SLL=-29.83 \\
 		\hline\hline
 		
 		\multirow{5}{*}{\tabincell{c}{Taylor \\($N$=128, SLL=-50 dB)}} & 3.96$\times 10^{-3}$ & $\chi$=0.101, SLL=-30.7&--&$\chi$=0.117, SLL=-32.43&$\chi$=0.101, SLL=-37.34 \\ \cline{2-6}			& 2.76$\times 10^{-3}$ & $\chi$=0.117, SLL=-34.9&--&$\chi$=0.148, SLL=-35.25&$\chi$=0.101, SLL=-37.34\\
 		\cline{2-6}
 		& 1.20$\times 10^{-3}$&--&$\chi$=0.101, SLL=-34.11&$\chi$=0.226, SLL=-41.71&$\chi$=0.117, SLL=-47.67 \\
 		\cline{2-6}
 		& 9.45$\times 10^{-4}$ & --&$\chi$=0.117, SLL=-35.70&$\chi$=0.258, SLL=-42.97&$\chi$=0.117, SLL=-47.67 \\
 		\cline{2-6}
 		& 4.36$\times 10^{-4}$ & --&$\chi$=0.133, SLL=-36.23&$\chi$=0.384, SLL=-44.30&$\chi$=0.117, SLL=-47.67 \\
 		\cline{2-6}
 		\hline\hline

 		\multicolumn{2}{|c||}{\tabincell{c}{Is antenna elements positions \\ optimization taken into account?}}& $\times$ & $\checkmark$&$\times$&$\checkmark$\\
 		\hline
 		
 	\end{tabular}
 \end{table*}
\section{Simulation Results}\label{SR}
 In this section, we are aimed at carrying out numerical experiments to evaluate the subarray synthesis performances of the proposed schemes in the cases when handling various array sizes and desired patterns with different SLLs, and compare them with competitive state-of-the-art subarray synthesis methods.

 First of all, two pattern matching indexes are introduced, the pattern matching error $\xi$ and the subarray synthesis rate $\chi$, which are formulated as follows:
 \begin{equation}\label{xi}
 	\xi =\frac{\int_{0}^{\pi/2} \vert\bar{F}(\theta)-F(\theta)\vert^2 d\theta}{\int_{0}^{\pi/2} \vert F(\theta)\vert^2 d\theta}
 \end{equation}
 \begin{equation}
	\chi = \frac{K}{N}
\end{equation}
where $F(\theta)$ is the achieved pattern and $K$ is the number of control points in the array, i.e., the number of subarrays.
\begin{figure}
	\centering
	\includegraphics[width=0.475\textwidth]{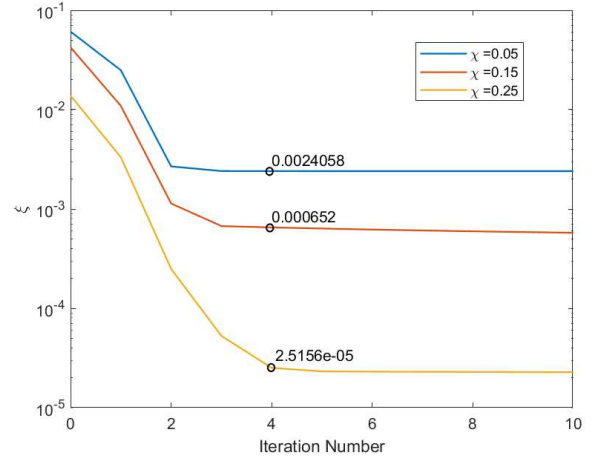}
	\caption{Curves of the effect of the number of iterations $\emph{Q}$ on $\xi$ under three values of $\chi=0.05$, $\chi=0.15$ and $\chi=0.25$ when the desired pattern is a Dolph-Tschebyscheff pattern with $N=20$ and SLL=-20 dB. }
	\label{Dolph(N=20)_iteration}
\end{figure}
For ease of elaboration and understanding, we use $\xi$ substiuting for $\varepsilon$ as the index for user-defined mode-2 , that is to say, the iteration termination conditions for user-defined mode-2 of proposed algorithms become an inequality about (\ref{xi}). Indeed, (\ref{xi}) can be regarded as the normalization of the restraints in (\ref{F}).
In order to demonstrate the generalization and effectiveness of the proposed methods, we consider three common desired patterns with different SLLs and array sizes as reference, which are Dolph-Tschebyscheff pattern, Taylor pattern and Flat-Top pattern. See Table \ref{all} for detailed parameter settings of the desired patterns, where all desired patterns are produced by uniformly half-wavelength spaced array. In addition, Table \ref{all} compares the proposed schemes with the state-of-the-art schemes under the subarray synthesis for mode-2 of different desired patterns. Specifically, it is based on the comparison of the achieved minimum subarray rate and SLL for the given user-defined matching error.

\subsection{Dolph-Tschebyscheff Pattern}

\begin{figure}
	\centering
	\includegraphics[width=0.475\textwidth]{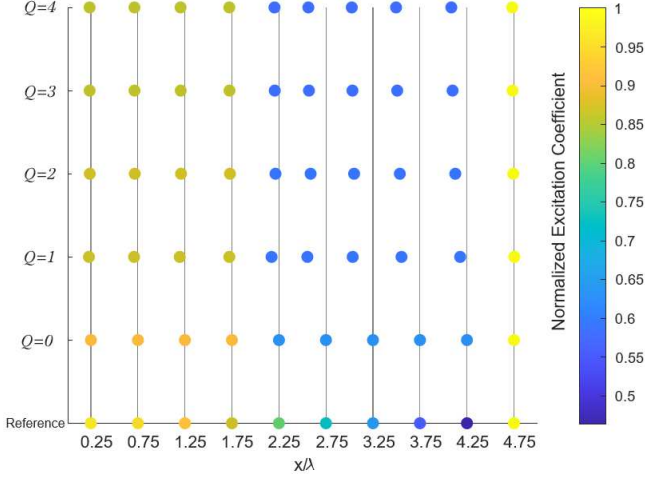}
	\caption{Changing process of the antenna elements positions and corresponding excitation coefficients of OGOMP-SS ($\chi=0.25$). Reference denotes the array parameters of Dolph-Tschebyscheff pattern with $N=20$ and SLL=-20 dB and $\emph{Q=0}$ means OMP-SS is adopted due to no iteration of OGOMP-SS.}
	\label{change_process}
\end{figure}

\begin{figure}
	\centering
	\includegraphics[width=0.475\textwidth]{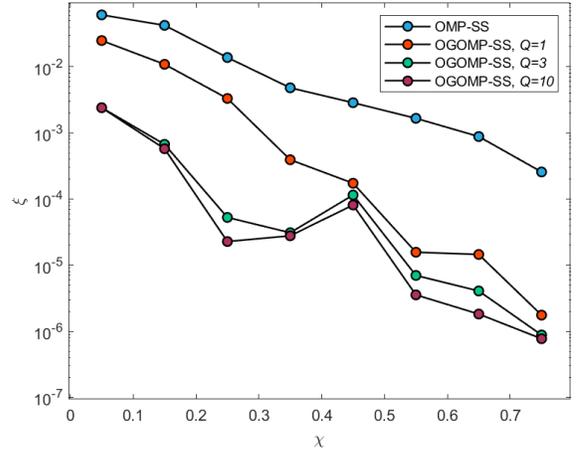}
	\caption{Pattern matching errors of OMP-SS and OGOMP-SS versus subarray rate $\chi$. The desired pattern is a Dolph-Tschebyscheff pattern with $N=20$ and SLL=-20 dB. }
\end{figure}

To validate the effectiveness of the proposed methods, we first evaluate the performances of the proposed schemes in the scenario where two Dolph-Tschebyscheff patterns with different parameters ($N$=20, SLL=-20 dB and $N$=100, SLL=-30 dB) are achieved.
Before achieving different patterns, it is of paramount significance to explore the impact of the iteration number $\emph{Q}$ in Algorithm \ref{off-omp} on the performance of the subarray synthesis. Towards this end, \figurename{3} exhibits the curves of synthesis performances of achieving the Dolph-Tschebyscheff pattern ($N$=20, SLL=-20 dB) with respect to the number of iterations $\emph{Q}$ in the cases when $\chi=0.05$, $\chi=0.15$ and $\chi=0.25$. From this, we can see that $\xi$ is in negative proportion to $\chi$ (i.e., the larger $\chi$, the smaller $\xi$) under the same number of iterations, and the performance curves generally start to converge after about 4 iterations. Especially, the iteration number $\emph{Q}=0$ means only OMP-SS is adopted due to no iteration of OGOMP-SS, that is, the antenna elements positions are not optimized. When $\emph{Q}>0$, the performance is greatly improved because the elements positions optimization are taken into account followed by the corresponding excitation coefficients updated. \figurename{4} shows the process of antenna elements positions and corresponding normalized excitation coefficients changing with iteration when adopting OGOMP-SS with $\chi=0.25$ and compares these array paramters with parameters obtained via OMP-SS (i.e., $\emph{Q}=0$) and that of the reference pattern. Since the linear antenna array is symmetrical, we only show the elements of the positive half axis of x. 
\begin{figure}
	\centering
	\includegraphics[width=0.405\textwidth]{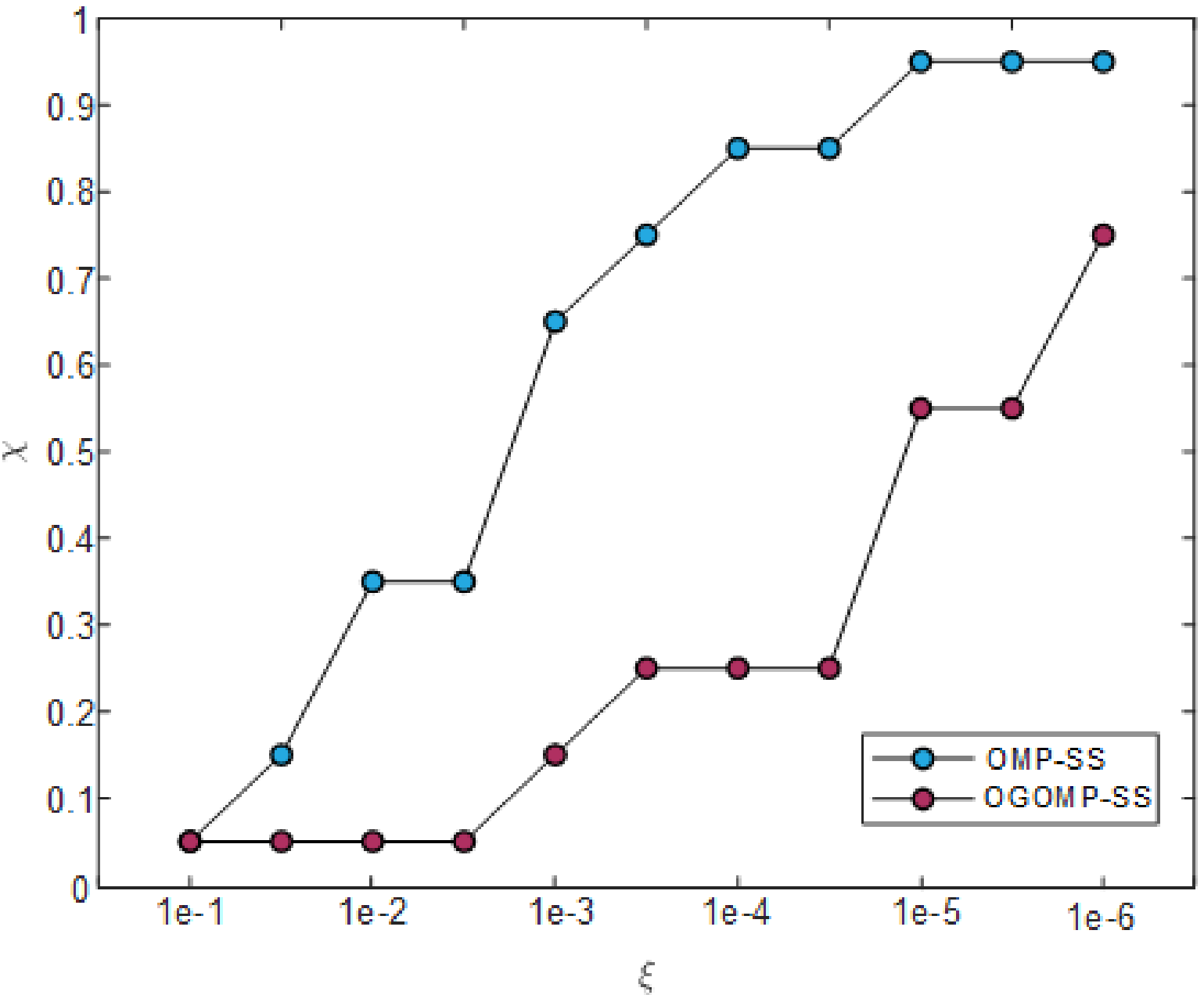}
	\caption{Minimum subarray rates of OMP-SS and OGOMP-SS versus pattern matching error $\xi$. The desired pattern is a Dolph-Tschebyscheff pattern with $N=20$ and SLL=-20 dB. }
\end{figure}

\begin{table}
	\caption{Array parameters corresponding to the desiredpattern and the achieved pattern }
	\label{excitation_20}
	\centering
	\begin{tabular}{ |c || c || c | c|c||c|  }
		\hline
		\multicolumn{2}{|c||}{\tabincell{c}{Dolph-Tschebyscheff Pattern \\($N=20$, SLL=-20 dB)}} &\multicolumn{2}{|c|}{\tabincell{c}{Achieved Pattern \\($\chi=0.25$, $\xi$=2.27e-5)}}\\
		\hline\hline
		\tabincell{c}{Location \\ ($d/\lambda$)}&\tabincell{c}{Normalized excitation\\ coefficient ($w$)}  &\tabincell{c}{Location \\ ($d/\lambda$)} &\tabincell{c}{Normalized excitation\\ coefficient ($w$)}  \\ \hline
		$\pm$0.2500	&0.9726&$\pm$0.2360 & 0.8583  \\ \hline
		$\pm$0.7500	&0.9546&$\pm$0.7133 & 0.8583  \\ \hline 
		$\pm$1.2500	&0.9193&$\pm$1.2043 & 0.8583 \\ \hline
		$\pm$1.7500	&0.8682&$\pm$1.7233 & 0.8583  \\ \hline
		$\pm$2.2500	&0.8034&$\pm$2.2036 & 0.5804  \\ \hline 
		$\pm$2.7500	&0.7274&$\pm$2.5621 & 0.5804  \\ \hline
		$\pm$3.2500	&0.6434&$\pm$3.0233 & 0.5804  \\ \hline
		$\pm$3.7500	&0.5544&$\pm$3.4923 & 0.5804  \\ \hline 
		$\pm$4.2500	&0.4639&$\pm$4.0806 & 0.5804  \\ \hline
		$\pm$4.7500	&1.0000 &$\pm$4.7306 & 1.0000  \\ \hline
	\end{tabular}
\end{table}

\begin{figure}[!htb]
	\centering
	\subfigure[Desired pattern and achieved patterns]{\includegraphics[width=0.5\textwidth]{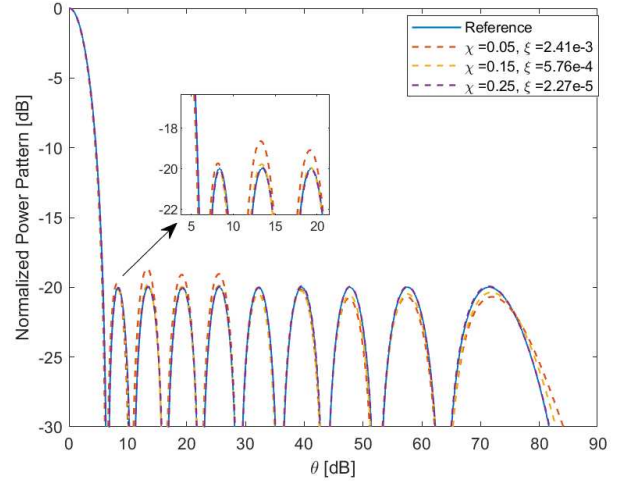}} \\ \subfigure[Excitation coefficients and elements positions corresponding to the patterns ]{\includegraphics[width=0.5\textwidth]{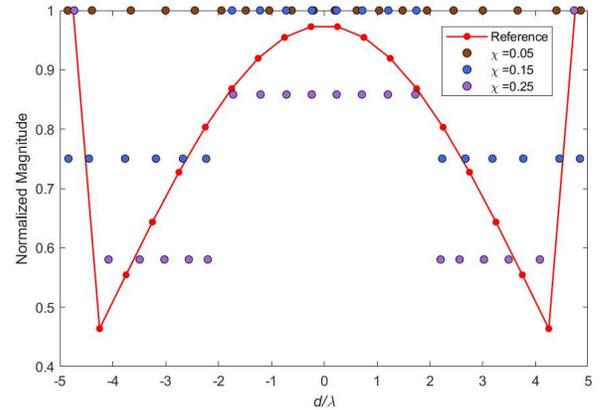}}
	\caption{Outcome of the subarray synthesis for a Dolph-Tschebyscheff pattern with $N$=20 and SLL=-20 dB in the cases when $\chi=0.05$, $\chi=0.15$ and $\chi=0.25$. }
	\label{Dolph_20}
\end{figure}
  Another significant simulation for evaluating the performance of OGOMP-SS in the scenario where we set $\emph{Q}$, $\chi$ to different values and OMP-SS as the baseline  is shown in \figurename{5}, therein, an interesting observation is that the performance curve of OGOMP-SS is not monotonic when the number of iterations is slightly large ($\xi$ at $\chi$ = 0.25 or 0.35 is amazingly smaller than $\xi$ at $\chi$ = 0.45, i.e., the change of the former  is greater than that of the latter  as the number of iterations increases), while the curve of OMP-SS in contrast is monotonic. This is due in large part to the potential of the non-uniform array because this does not happen in the uniform array according to the simulation result of OMP-SS. There are two meanings: i) ignoring the optimization of antenna elements positions, the subarray synthesis performance will increase as the subarray rate $\chi$ increases and the performance is optimal when the number of subarrays is equal to that of antennas, ii)  there is a possibility that the  subarray synthesis performance of low subarray rate is better than that of high subarray rate in the non-uniform array. To sum up, figures 1 to 3 demonstrate the significance of the optimization of antenna elements positions on the one hand and reveal that good performance can also be achieved by optimizing the antenna position in the case of low subarray rate on the other hand. In addition, it should be specially noted that $\emph{Q}=10$ will be set for the next simulations.

  It is noteworthy that our proposed schemes not only support the subarray synthesis for the given subarray rate $\chi$, but also can directly carry out the subarray synthesis under the user-defined accuracy requirements (the pattern matching error $\xi$), i.e., the user-defined mode-2. Thus, \figurename{6} illustrates the curves of the minimum subarray rate $\chi$ of the mode-2 of OMP-SS and OGOMP-SS schemes under the condition of meeting different pattern matching errors $\xi \in [1e-6,1e-1]$, from which we can see that OGOMP-SS can satisfy the user-defined error $\xi$ with less subarray rate $\chi$ compared with OMP-SS.
  
\figurename{7} exhibits the results achieved for the Dolph-Tschebyscheff pattern ($N$=20, SLL=-20 dB) in the cases when $\chi=0.05$, $\chi=0.15$ and $\chi=0.25$, where \figurename{7} (a) displays the desired pattern and achieved patterns when $\theta$ ranges from $0^{\circ}$ to $90^{\circ}$ and \figurename{7} (b) visualizes the elements positions and their corresponding normalized excitation coefficients for the four patterns. In addition, Table \ref{excitation_20} expounds the position of each antenna element and its excitation coefficient when achieving the Dolph-Tschebyscheff pattern with $N=20$ and SLL=-20 dB via OGOMP-SS with $\chi=0.25$ .

As for a further simulation, we choose to expand the aperture of the array with more challenging SLL requirement, i.e., the Dolph-Tschebyscheff pattern with $N=100$ and SLL=-30 dB is used for reference.

\begin{figure}
	\centering
	\includegraphics[width=0.475\textwidth]{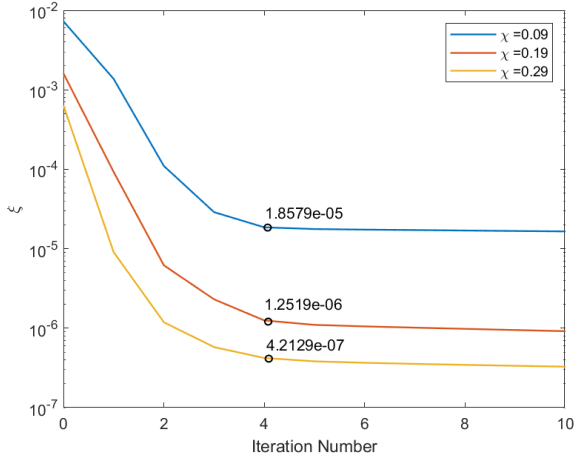}
	\caption{Curves of the effect of the number of iterations $\emph{Q}$ on $\xi$ under three values of $\chi=0.09$, $\chi=0.19$ and $\chi=0.29$ when the desired pattern is Dolph-Tschebyscheff pattern with $N=100$ and SLL=-30 dB. }
	\label{Dolph(N=100)_iteration}
\end{figure}
\begin{figure}
	\centering
	\includegraphics[width=0.475\textwidth]{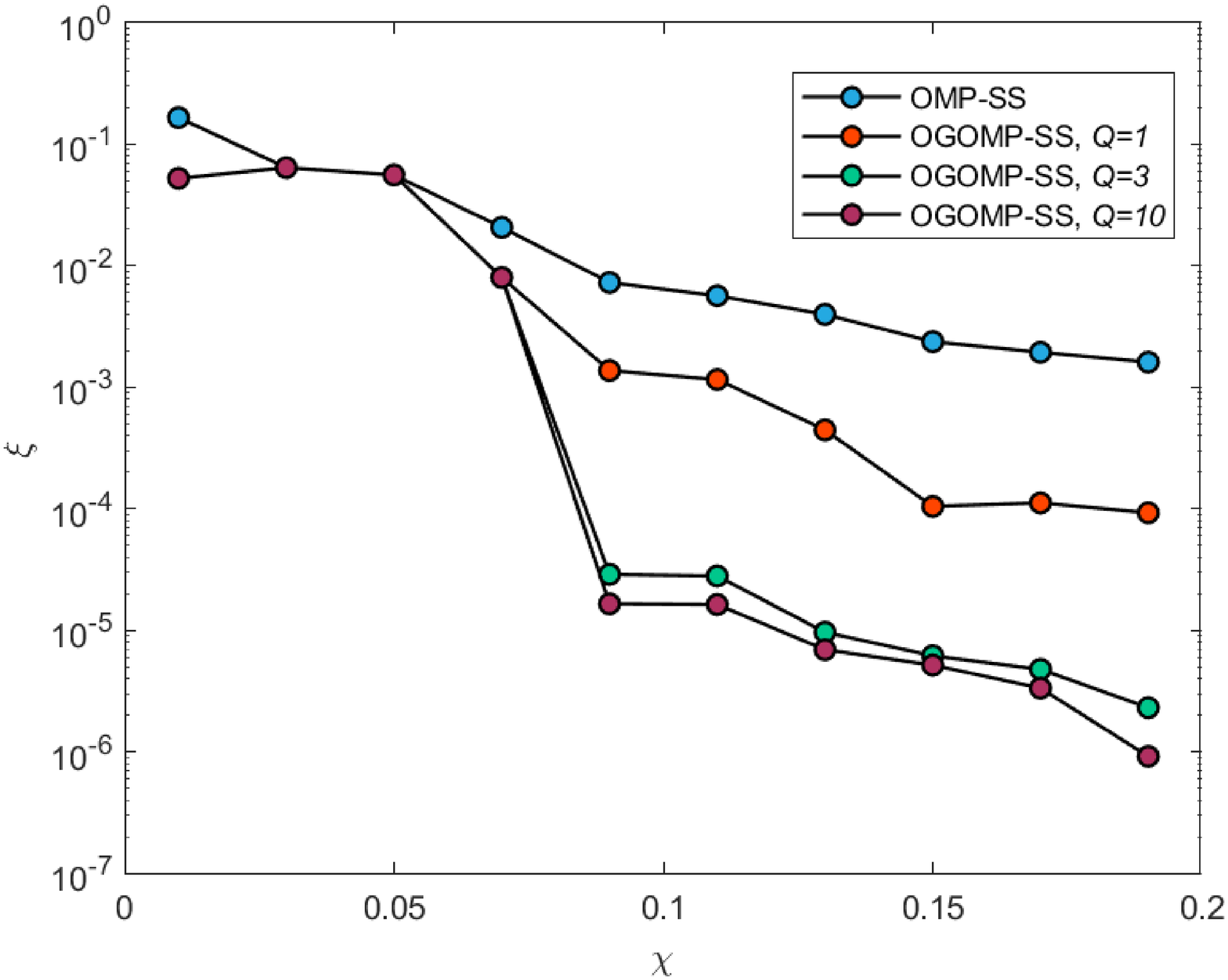}
	\caption{Pattern matching errors of OMP-SS and OGOMP-SS versus subarray rate $\chi$. The desired pattern is a Dolph-Tschebyscheff pattern with $N=100$ and SLL=-30 dB. }
\end{figure}
\begin{figure}
	\centering
	\includegraphics[width=0.405\textwidth]{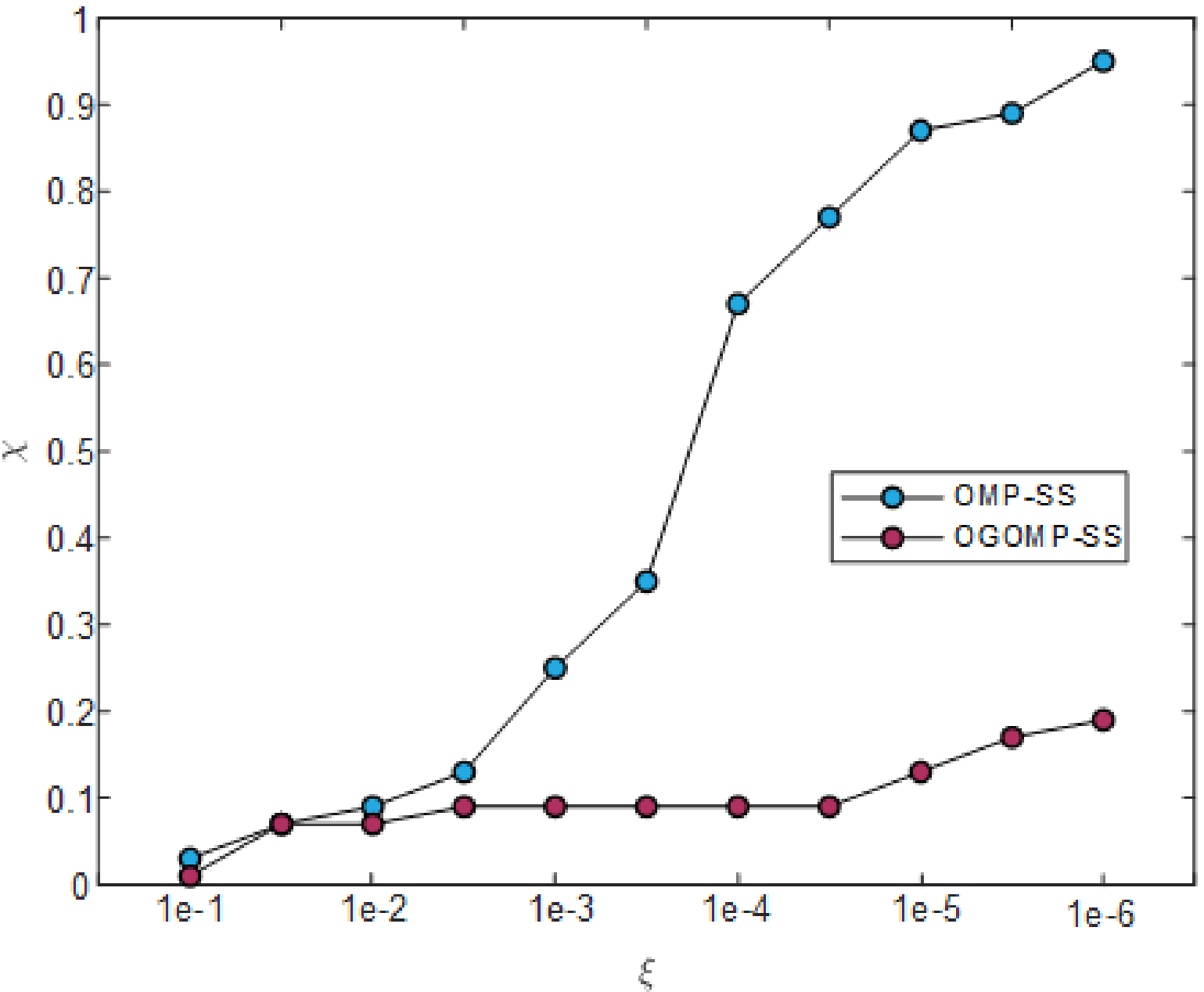}
	\caption{Minimum subarray rates of OMP-SS and OGOMP-SS versus pattern matching error $\xi$. The desired pattern is a Dolph-Tschebyscheff pattern with $N=100$ and SLL=-30 dB. }
\end{figure}

\begin{figure}[!htb]
	\centering
	\subfigure[Desired pattern and achieved patterns]{\includegraphics[width=0.5\textwidth]{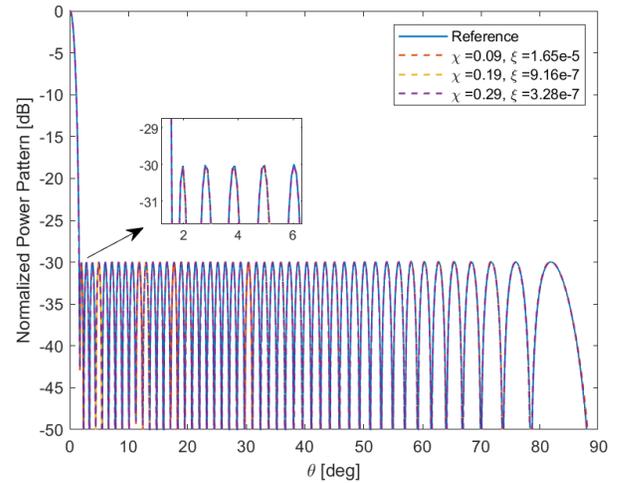}} \\ \subfigure[Excitation coefficients and elements positions corresponding to the patterns ]{\includegraphics[width=0.5\textwidth]{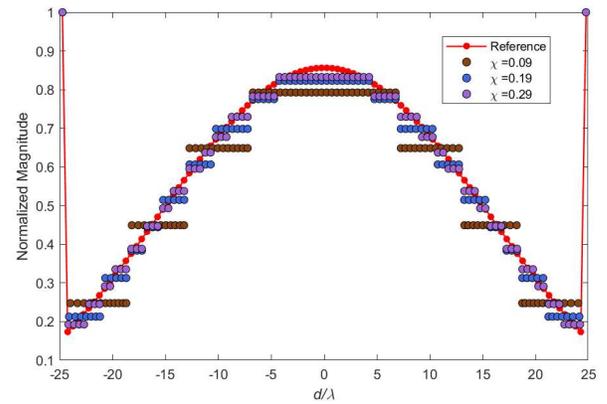}}
	\caption{Outcome of the subarray synthesis for a Dolph-Tschebyscheff pattern with $N$=100 and SLL=-30 dB in the cases when $\chi=0.09$, $\chi=0.19$ and $\chi=0.29$. }
	\label{Dolph_100}
\end{figure}
\figurename{8} shows the curves of synthesis performances of achieving the Dolph-Tschebyscheff pattern ($N$=100, SLL=-30 dB) with respect to the number of iterations $\emph{Q}$ in the cases when $\chi=0.09$, $\chi=0.19$ and $\chi=0.29$, From which we can see that similar to the result of achieving the Dolph-Tschebyscheff pattern with $N=20$ and SLL=-20 dB, the performance curves start to converge after about 4 iterations. On the one hand, \figurename{9} exhibits the pattern matching errors of OMP-SS and OGOMP-SS with different iterations versus the subarray rate, which is used for user-defined mode-1. 
\begin{figure}
	\centering
	\includegraphics[width=0.475\textwidth]{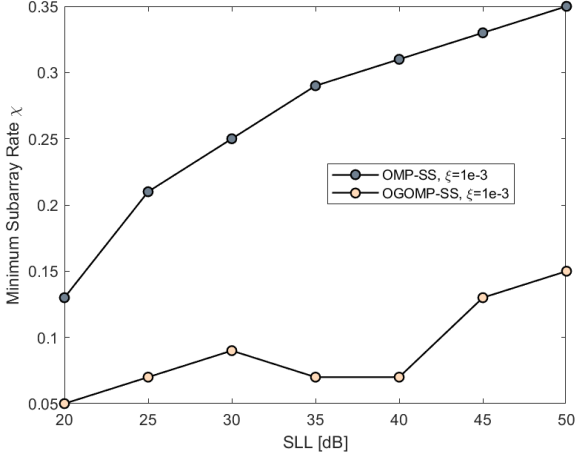}
	\caption{Curves of minimum subarray rates required of OMP-SS and OGOMP-SS ($\emph{Q}$=10) when achieving the Dolph-Tschebyscheff pattern with $N=100$ and SLL varying 20 to 50 under the user-defined error $\xi=1\times 10^{-3}$.}
	\label{Dolph_SLLvar}
\end{figure}

\begin{figure}
	\centering
	\includegraphics[width=0.475\textwidth]{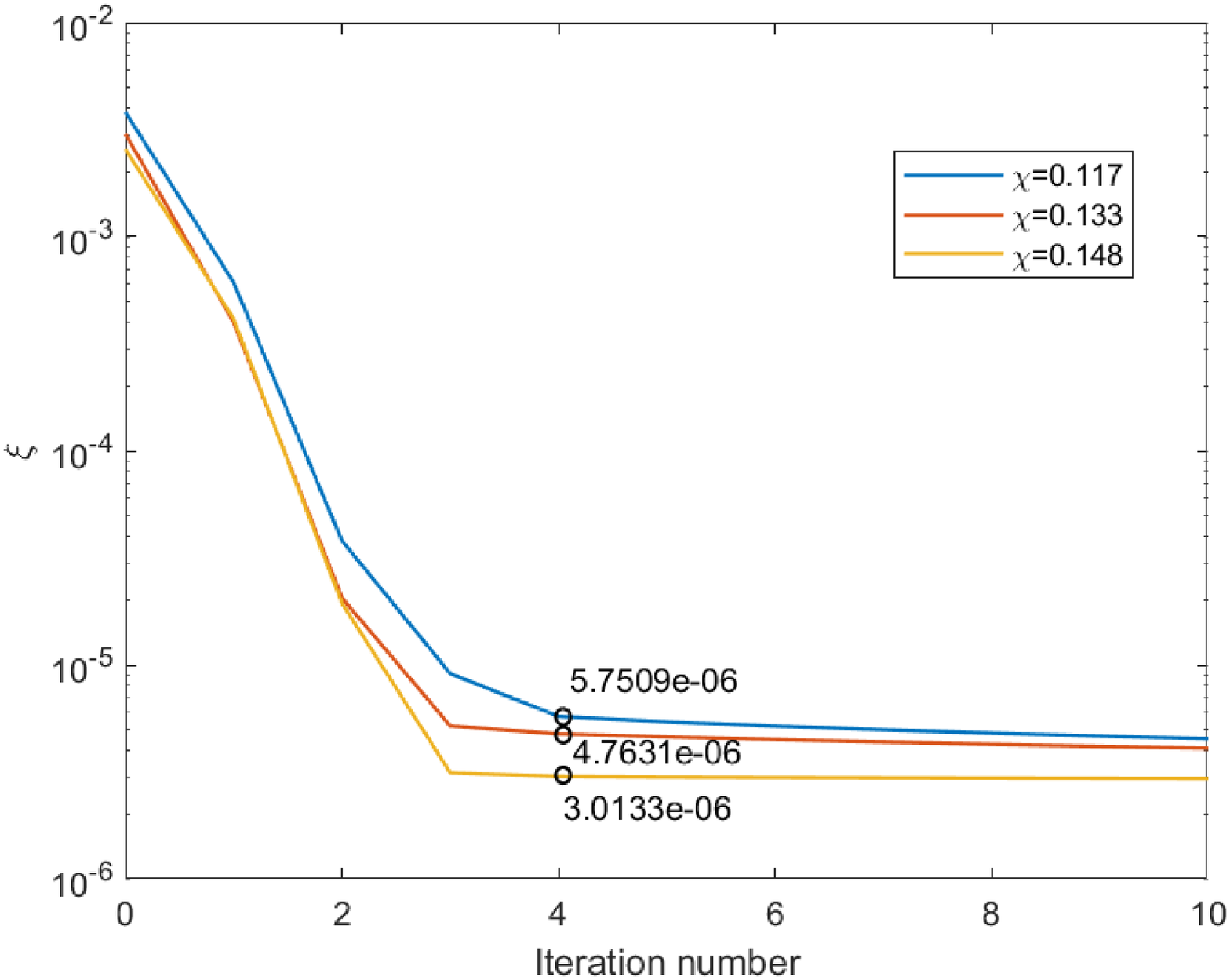}
	\caption{Curves of the effect of the number of iterations $\emph{Q}$ on $\xi$ under three values of $\chi=0.117$, $\chi=0.133$ and $\chi=0.148$ when the desired pattern is Taylor pattern with $N=128$ and SLL=-50 dB. }
	\label{taylor_iteration}
\end{figure}
On the other hand, \figurename{10} illustrates the curves of minimum subarray rate required to meet different matching errors for user-defined mode-2, wherein, the excellent performance of OGOMP-SS in minimizing subarray rate with high matching accuracy is revealed in comparison with OMP-SS. Furthermore, \figurename{11} exhibits the outcome achieved for the Dolph-Tschebyscheff pattern ($N$=100, SLL=-30 dB) in the cases when $\chi=0.09$, $\chi=0.19$ and $\chi=0.29$, where \figurename{11} (a) displays the desired pattern and achieved patterns when $\theta$ ranges from $0^{\circ}$ to $90^{\circ}$. The achieved patterns are very close to the desired pattern due to the pattern matching error $\xi$ of each achieved pattern is small enough. As can be seen from the geometry of \figurename{11} (b), visualizing the elements positions and their corresponding normalized excitation coefficients for the four patterns, the distribution of the excitation coefficients of each realized pattern is similar to that of the desired pattern.

For the purpose of exploring the influence of SLL of the desired pattern on the subarray synthesis performance, we draw the curves of minimum subarray rates required of OMP-SS and OGOMP-SS ($\emph{Q}=10$) when achieving the Dolph-Tschebyscheff pattern with $N=100$ and SLL varying 20 to 50 under the user-defined error $\xi=1\times 10^{-3}$ in \figurename{12}. 

\subsection{Taylor Pattern}
\begin{figure}
	\centering
	\includegraphics[width=0.469\textwidth]{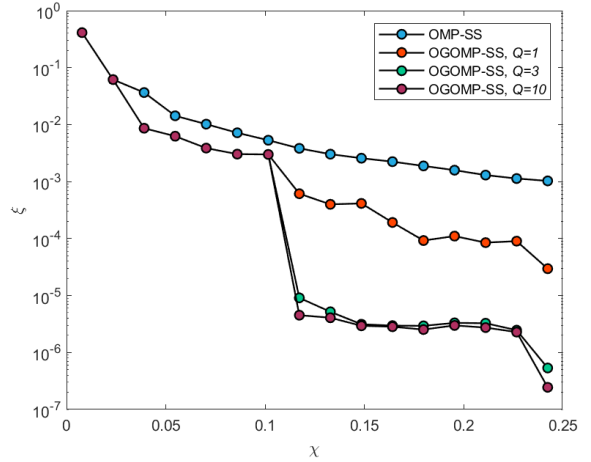}
	\caption{Pattern matching errors of OMP-SS and OGOMP-SS versus subarray rate $\chi$. The desired pattern is a Taylor pattern with $N=128$ and SLL=-50 dB. }
	
\end{figure}
\begin{figure}
	\centering
	\includegraphics[width=0.405\textwidth]{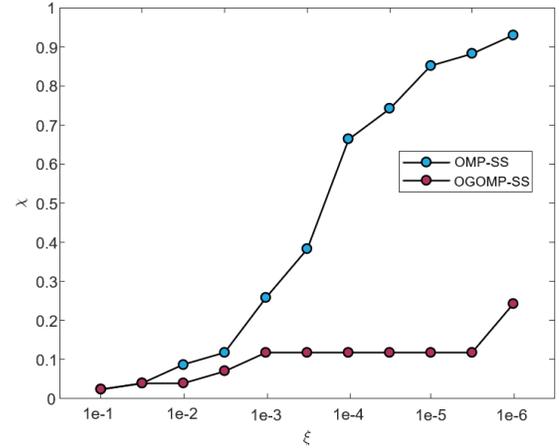}
	\caption{Minimum subarray rates of OMP-SS and OGOMP-SS versus pattern matching error $\xi$. The desired pattern is a Taylor pattern with $N=128$ and SLL=-50 dB. }
\end{figure}
\begin{figure}[!htb]
	\centering
	\subfigure[Desired pattern and achieved patterns]{\includegraphics[width=0.5\textwidth]{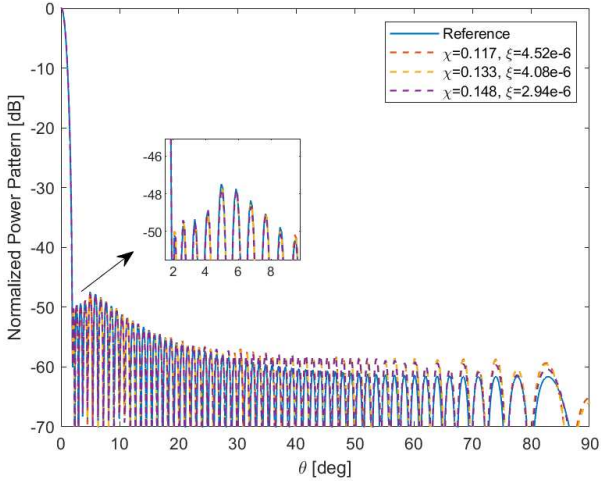}} \\ \subfigure[Excitation coefficients and elements positions corresponding to the patterns ]{\includegraphics[width=0.5\textwidth]{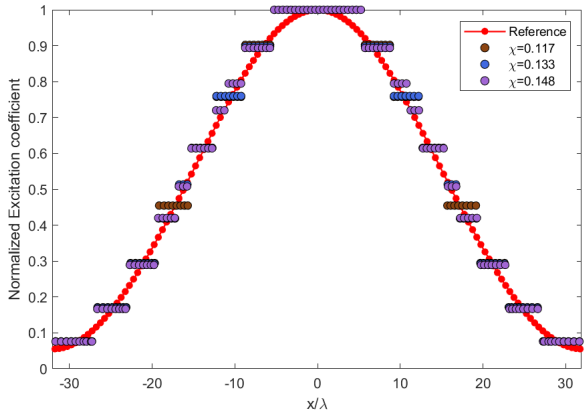}}
	\caption{Outcome of the subarray synthesis for a Taylor pattern with $N$=128 and SLL=-50 dB in the cases when $\chi=0.117$, $\chi=0.133$ and $\chi=0.148$. }
	\label{talor}
\end{figure}
In order to demonstrate the generalization of our proposed schemes, a discrete Taylor pattern with $\bar{n}=5$ (number of constant level sidelobes), $N=128$ and SLL=-50 dB is achieved, which is also considered in \cite{sub-cs3,sub-cs4}. \figurename{13} displays the curves of synthesis performances of achieving the Taylor pattern with respect to the number of iterations $\emph{Q}$ in the cases when $\chi=0.117$, $\chi=0.133$ and $\chi=0.148$, in like wise, $\emph{Q}=0$ represents OMP-SS scheme. In addtion, the curves of different $\chi$ start to converge after about 4 iterations. There exhibits the pattern matching errors of OMP-SS and OGOMP-SS with different iterations in the cases when user-defined $\chi$ varies from 0.008 to 0.242 in \figurename{14}, wherein, the performance improves greatly when the number of iterations changes from 0 to 3. In \figurename{15}, the curves of minimum subarray rates of OMP-SS and OGOMP-SS versus user-defined $\xi$ are drawn to illustrate OGOMP-SS can satisfy high matching accuracy at very low subarray rates in achieving the Taylor pattern. 

Moreover, \figurename{16} exhibits the subarray synthesis outcome for achieving the Taylor pattern ($N$=128, SLL=-50 dB) in the cases when $\chi=0.117$, $\chi=0.133$ and $\chi=0.148$, in which \figurename{16} (a) displays the desired pattern and achieved patterns when $\theta$ ranges from $0^{\circ}$ to $90^{\circ}$ and \figurename{16} (b) visualizes the elements positions and their corresponding normalized excitation coefficients for the four patterns. In \figurename{17}, we draw the curves of minimum subarray rates required of OMP-SS and OGOMP-SS ($\emph{Q}=10$) when achieving the Taylor pattern with $N=128$ and SLL varying 20 to 50 under the user-defined error $\xi=1\times 10^{-3}$ to reveal the influence of SLL of the desired pattern on the subarray synthesis performance.

 \begin{figure}
 	\centering
 	\includegraphics[width=0.475\textwidth]{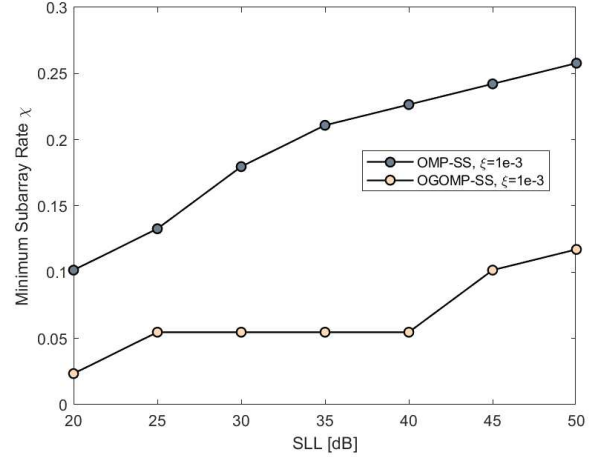}
 	\caption{Curves of minimum subarray rates required of OMP-SS and OGOMP-SS ($\emph{Q}$=10) when achieving the Taylor pattern with $N=128$ and SLL varying 20 to 50 under the user-defined error $\xi=1\times 10^{-3}$.}
 	\label{Taylor_SLLvar}
 \end{figure}
 
\section{Conclusions}\label{Con}
In this paper, we first propose an on-grid CS framework developed by OMP for the subarray synthesis to attain the subarray layout (the number of subarrays, the number of elements per subarray) and excitation coefficients, wherein, there are two user-defined modes provided for different synthesis scenarios. Furthermore, an off-grid OMP for the subarray synthesis is proposed to optimize the antenna elements positions, subarray layout and excitation coefficients simultaneously, which achieves excellent synthesis performance compared with other competitive state-of-the-art subarry synthesis methods. In addition, simulation results exhibit the synthesis performance of two user-defined modes of the proposed schemes in the cases when achieving different patterns.

\bibliographystyle{IEEEtran}
\bibliography{reference.bib}

\vspace{12pt}

\end{document}